\documentclass[jmse,article,accept,moreauthors,pdftex]{Definitions/mdpi} 
\usepackage{appendix}
\usepackage{caption}
\usepackage{subcaption}
\firstpage{1} 
\pubvolume{J. Mar. Sci. Eng.}
\issuenum{8}
\articlenumber{127}
\pubyear{2020}
\copyrightyear{2019}
\history{Received: 18 December 2019 / Revised: 29 January 2020 / Accepted: 10 February 2020 / Published: 17 February 2020}

\Title{Improved VIV response prediction using adaptive parameters and data clustering}

\Author{Jie Wu $^{1}$, Decao Yin $^{1}$*, Halvor Lie $^{1}$, Signe Riemer-Sørensen $^{2}$, Svein Sævik $^{3}$ and Michael Triantafyllou $^{4}$}

\AuthorNames{Firstname Lastname, Firstname Lastname and Firstname Lastname}

\address{%
$^{1}$ \quad SINTEF Ocean; P.O. Box 4762 Torgarden, 7465 Trondheim, Norway; jie.wu@sintef.no (J.W.); decao.yin@sintef.no (D.Y.); halvor.lie@sintef.no (H.L.)\\
$^{2}$ \quad SINTEF Digital; P.O. Box 124 Blindern, 0314 Oslo, Norway; signe.riemer-sorensen@sintef.no (S. R.-S.)\\
$^{3}$ \quad NTNU, Faculty of Engineering, Department of Marine Technology;  7491 Trondheim, Norway; svein.savik@ntnu.no (S. S.)\\
$^{4}$ \quad MIT, Department of Mechanical Engineering, Cambridge, Massachusetts 02139, USA;
mistetri@mit.edu (M. T.)}
\corres{Correspondence: decao.yin@sintef.no (D.Y.); Tel.: +47 9199 7166}

\abstract{Slender marine structures such as deep-water riser systems are continuously exposed to currents leading to vortex-induced vibrations (VIV) of the structure. This may result in amplified drag loads and fast accumulation of fatigue damage. Consequently, accurate prediction of VIV responses is of great importance for the safe design and operation of marine risers. Model tests with elastic pipes have shown that VIV responses are influenced by many structural and hydrodynamic parameters, which have not been fully modelled in present frequency domain VIV prediction tools. Traditionally, predictions have been computed using a single set of hydrodynamic parameters, often leading to inconsistent prediction accuracy when compared with observed field measurements and experimental data. Hence, it is necessary to implement a high safety factor of 10 - 20 in the riser design, which increases development cost and adds extra constraints in the field operation. One way to compensate for the simplifications in the mathematical prediction model is to apply adaptive parameters to describe different riser responses. The objective of this work is to demonstrate a new method to improve the prediction consistency and accuracy by applying adaptive hydrodynamic parameters. In the present work, a four-step approach has been proposed: First, the measured VIV response will be analysed to identify key parameters to represent the response characteristics. These parameters will be grouped by using data clustering algorithms. Secondly, optimal hydrodynamic parameters will be identified for each data group by optimisation against measured data. Thirdly, the VIV response using the obtained parameters will be calculated and the prediction accuracy evaluated. Last but not the least, classification algorithms will be applied to determine the correct hydrodynamic parameters to be used for new cases. An iteration of the previous steps may be needed if the prediction accuracy of the new case is not satisfactory. This concept has been demonstrated with examples from experimental data.}

\keyword{Vortex-induced vibrations; model test; hydrodynamics; machine learning; data clustering; data classification}

\begin{document}

\section{Introduction}
Risers and pipelines are important structural elements in offshore oil and gas exploration and development. Vortex-induced vibrations (VIV) cause cyclic loads and amplified drag force on the cylindrical structures. Consequently, severe fatigue damage may be accumulated and mechanical failure can occur. Therefore, in the design and operation of marine risers, it is important to understand the global VIV responses of deepwater risers and marine pipelines, and get accurate prediction of the responses.

Semi-empirical VIV prediction tools such as VIVANA \cite{Larsen2009}, Shear7 \cite{shear7} and VIVA \cite{viva99} are tools commonly used by the industry. These prediction tools consist of a structural model and a hydrodynamic load model. The structural model may be based on normal or complex mode superposition or use finite elements with coupled equations. The hydrodynamic load model is based on the ``strip theory''. The hydrodynamic force and frequency depends on local flow conditions, response amplitude and structural geometry. The load model and the empirical parameters are developed from model tests using rigid cylinders and flexible pipes. 

The hydrodynamic parameters required by the prediction tools are normally generalised from rigid cylinder forced motion tests. In such tests, a rigid cylinder is towed through the water at a constant speed ($U$) and forced to oscillate under designed motion amplitudes ($A$) and frequencies ($f_{osc}$), which are approximating the actual trajectories of the cross-section of a flexible pipe. The hydrodynamic coefficients are derived from the force measurement of the rigid cylinder section \cite{gopthesis,aronthenthesis}. Despite the efforts to obtain hydrodynamic data from realistic cylinder motions \cite{dahlthesis,zheng2011,yinthesis,aglenthesis}, the hydrodynamic parameters used in the present prediction tools are still based on tests with one-dimensional cross-flow (CF) harmonic motions, neglecting in-line (IL) vibrations that are also present. Often these tests are carried out in conditions characterised by sub-critical Reynolds numbers, while the riser in the field will experience flow condition in the critical or super-critical flow regimes.

The effects of variations in the Reynolds number were studied on a full scale pipe section by \cite{otc16341,lie2013,yin2018b}. It was found that the hydrodynamic coefficient is sensitive to the Reynolds number. In addition, the surface roughness ratio was  also found to be an important parameter. These effects have not been accounted for in the present VIV prediction tools.

VIV tests with elastic pipe/riser models have mainly focused on global responses \cite{trim2005,otc8071,Vandiver2009,nielsen2002,lie2007,otc8700,chaplin2005} where bending strain, curvature and/or acceleration are typically measured at discrete locations along the test model under uniform and shear current conditions. The measured cross-section motion trajectories found from such tests are  far from harmonic \cite{yinthesis}. The hydrodynamic force coefficients along the test model can be estimated by an inverse analysis method \cite{wu2009, wu2010}. This method basically uses measured response data to calculate the external VIV loads by assuming that the mechanical properties (stiffness, mass, etc) are known. The hydrodynamic coefficients estimated by inverse analysis also show deviation from results of rigid cylinder tests. In such tests, small scale models are normally used, which means that the Reynolds number is relatively
low.

Inconsistent prediction accuracy is seen when benchmarked against elastic cylinder tests in ideal laboratory conditions, where tests are carried out under constant and uni-directional flow  with densely spaced and accurate measurements \cite{voie2017}. Even larger discrepancies have been seen when compared to full scale field measurement data \cite{Camp2010}. Part of the reasons for these discrepancies are over-simplifications in the hydrodynamic load model and the associated parameters applied in the prediction tools.

The objective of this work is to introduce a new method to improve the prediction consistency and accuracy by applying adaptive hydrodynamic parameters. In the present study, the complexity of VIV responses is first explained with typical model test examples in Sec. \ref{sec:characteristics}. The limitations of the present VIV prediction tools with built-in empirical parameters is discussed in Sec. \ref{sec:limitations}. This provides the background and reasons for using adaptive parameters to compensate for the simplifications in the numerical models. The concept of adaptive parameters is then discussed in Sec. \ref{sec:adaptive} with examples to demonstrate the possible improvement in VIV predictions. Machine Learning (ML) has been applied where  data clustering algorithms have been used to group the response data from various model tests and where each group is associated with a set of parameters. Classification algorithms have been used to select the optimum parameters for the new data. Potential applications to improve riser design practice and field monitoring are discussed in Sec. \ref{sec:adaptive}.

\section{Characteristics of VIV responses} \label{sec:characteristics}
It is known that the current profile can have significant influence on VIV responses of deep water riser systems. There is a lack of prediction models for describing VIV responses subjected to three-dimensional (3D) current \cite{Riemer2019}. Therefore, in the present design practice, a uni-directional (2D) current profile is assumed. Sheared or uniform 2D currents are normally simulated in model tests. If the current profile is linearly sheared, the excitation region is well defined in the high current speed area. If the shear profile is similar to the Gulf of Mexico profile \cite{bourguet2013}, the excitation region is distributed over several locations along the pipe length. Since the sheared current profile is more realistic for deep-water riser systems, five different VIV model tests with sheared current profiles were selected to be further investigated in the present work, with focus on the CF response.

\subsection{Influence of travelling speed and bending stiffness on VIV responses}
The responses of a elastic pipe can be influenced by the longitudinal travelling speed of the VIV response. \cite{wu2018}. 
 
The theoretical travel speed of oscillations of an infinitely long tensioned string and an un-tensioned beam are calculated by the following equations:

Tensioned string:
\begin{equation}
    C_s=\sqrt{\frac{T}{m}} 
\end{equation}
							
Un-tensioned beam:
\begin{equation}
    C_{b,CF/IL}=\sqrt[4]{\frac{\omega^2_{CF/IL} EI}{m}}
\end{equation}
where, $m$ is the mass per unit length; $EI$ is the bending stiffness; $T$ is the tension and $\omega$ is the response frequency in IL and CF directions.

For a tension-dominated string, the travel speed will be independent of the response frequency. This means that the wave speeds of responses in IL and CF directions are the same. For a beam with controlled bending stiffness, the longitudinal travelling speed depends on the response frequency. This means the CF response at the primary frequency $\omega$ will be travelling along the pipe with a different speed than the IL response. Thus, the motion trajectories will constantly vary along the pipe. This will affect CF VIV response both at the primary frequency $\omega$ 
and the third order frequency component $3\times \omega$ together with other factors. \cite{wu2018}

One way to quantify the importance of the bending stiffness is to evaluate the natural frequencies of the pipe model. Elastic pipe models in VIV tests could be simplified as tensioned elastic beams. Both tension and bending stiffness contribute to the total stiffness. The natural frequencies of a tensioned beam are given by the following equations: 

For a tensioned string:
\begin{equation}
    f_{n,s}=\frac{n}{2L}\sqrt{\frac{T}{m}}	
\end{equation}

For an un-tensioned beam: 
\begin{equation}
    f_{n,b}=\frac{n^2\pi}{2L^2}\sqrt{\frac{EI}{m}}
\end{equation}
						
For a tensioned beam:
\begin{equation}
    f_{n,tot}=\sqrt{f^2_{n,s}+f^2_{n,b}}
\end{equation}
where, $L$ is the pipe length; $n$ is the mode number.

The relative magnitude of the bending stiffness can be quantified by the frequency ratio 
\begin{equation}
F=f_{n,b}/f_{n,tot}
\end{equation}
The higher value of this parameter, the stronger the influence of the bending stiffness will be. Further, the travelling speed will  be increasingly different in the IL and CF directions.  

\subsection{Selected elastic pipe model tests}
Performing experiments on a scaled elastic riser or on pipeline models is a common approach to study the global response. The measurements are used to validate various numerical VIV prediction tools. In this section, several representative model tests of elastic bare riser models are reviewed, and the physical properties are summarized in Table \ref{tab:mt}. A description of the test setup can be found in the Appendix. 

\begin{table}[H]
\caption{Summary of physical properties of the riser/pipe model of various elastic model VIV tests.} \label{tab:mt}
\centering
\begin{tabular}{lccccc}
\toprule
\textbf{Test}	& Length 	& Diameter  & Mass ratio  & Bending stiffness  & Mean tension \\
    & $ L (m)$  & $D (m)$ & $m^* $  & $EI (Nm^2)$ & $\bar{T} (N)$ \\
\midrule
NDP high mode test	&	38&	0.027&	1.632&	37.2&	4000\\
Shell high mode test (pipe 2) 	&	38&	0.030&	1.54&	572.3& 4000	\\
Hanøytangen		&	90& 0.030	& 3.13	& 3639	&	3700	\\
ExxonMobil rotating rig test		&	9.63&	0.02& 2.17	&	135.4&	700	\\
Miami II test 			&	152.4 &	0.0363 &	1.268&	613 &	3225	\\
\bottomrule
\end{tabular}
\end{table}

\subsection{Characteristics of VIV responses of an elastic pipe in sheared current}
\begin{figure}
\centering
\includegraphics[width=0.95\textwidth]{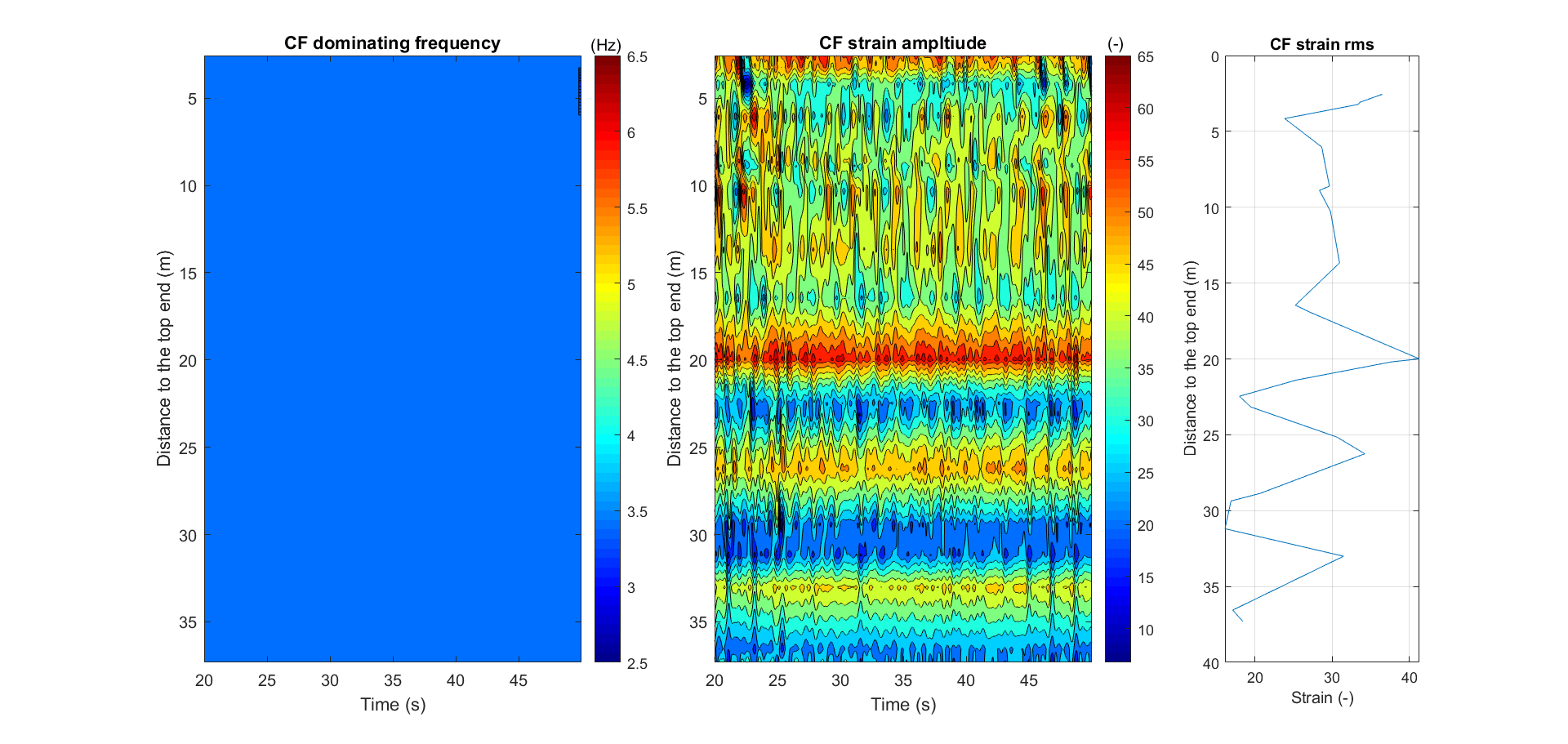}
         \caption{Single-frequency response, low mode ($\sim 5$), tension dominating, NDP high mode test case 2350 \cite{trim2005}.}
         \label{fig:single_freq}
\end{figure}

\begin{figure}
\centering
\includegraphics[width=0.95\textwidth]{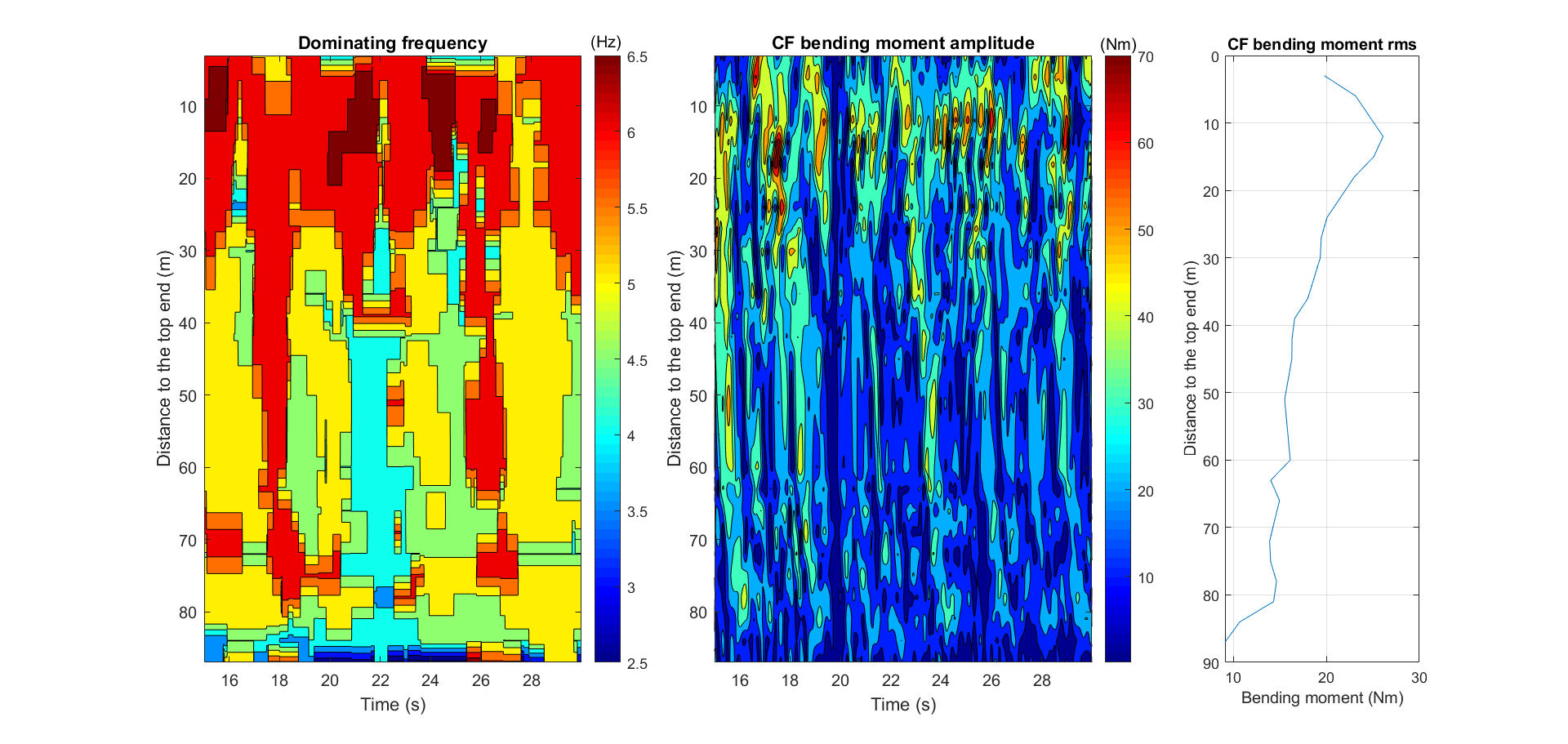}
         \caption{Multi-frequency response, high mode ($\sim 25$), bending stiffness dominating, Hanøytangen test case 0056          \cite{otc8071}.}
         \label{fig:multi_freq}
\end{figure}

The shedding frequency is given by $f_s= StU/D$, where St is the Strouhal number,  $D$ is the pipe outer diameter. In a sheared flow, the shedding frequency may vary along the length of the pipe. In addition, the response frequency and mode are known to vary in time even under a constant flow speed due to the flow instability. The variation of the dominating response frequency in space and time can be extracted from wavelet analysis of the measured response signals along the length of the test pipe \cite{OMAE2012-83092}. The frequency composition can be different for a tension dominated pipe vibrating at a low mode ($\sim 5$), as shown in Fig. \ref{fig:single_freq} or bending stiffness dominated pipe responding at a high mode ($\sim 25$), as shown in Fig. \ref{fig:multi_freq}. The response is dominated by a single frequency ($\approx 3.5~Hz$) for the former, even when subjected to a sheared current. However, the frequency variation is much more complicated for high mode VIV responses, as shown in Fig. \ref{fig:multi_freq}. Higher frequencies are observed at the high flow speed region at the top of the riser, and lower frequencies close to the bottom where the flow speed decreases to zero (``space sharing'' \cite{Larsen2009}), e.g., 15 – 17 s. ``Time sharing'' \cite{SSthesis} of the frequency is also observed. The experimental results show that the response frequency varies along the length of the pipe. If well synchronised, fewer frequencies tend to dominate longer sections of the pipe. The response will also be stable with higher motion amplitudes. On the other hand, the response will be controlled by the local shedding process when the synchronisation breaks down, which leads to more frequencies and chaotic responses with lower motion amplitudes. 

Such behaviour is related to the bending stiffness of the structure, which can affect the response wave propagation speed and motion orbits. It is also related to the presence of the travelling waves. The higher mode, the more travelling wave response will be present, as discussed earlier. In addition, large tension variation has been observed in some of the high mode VIV response tests, which may also contribute to the instability of the response.

The model tests with long elastic cylinder models cover a large variation of riser dimensions, cross-sectional properties, boundary conditions, and flow conditions. This calls for an improved understanding the VIV behaviours of different riser configurations and numerical tools with increased robustness and specialised treatment of the response characteristics \cite{vivbp2016}.

\begin{figure}
\centering
\includegraphics[width=0.8\textwidth]{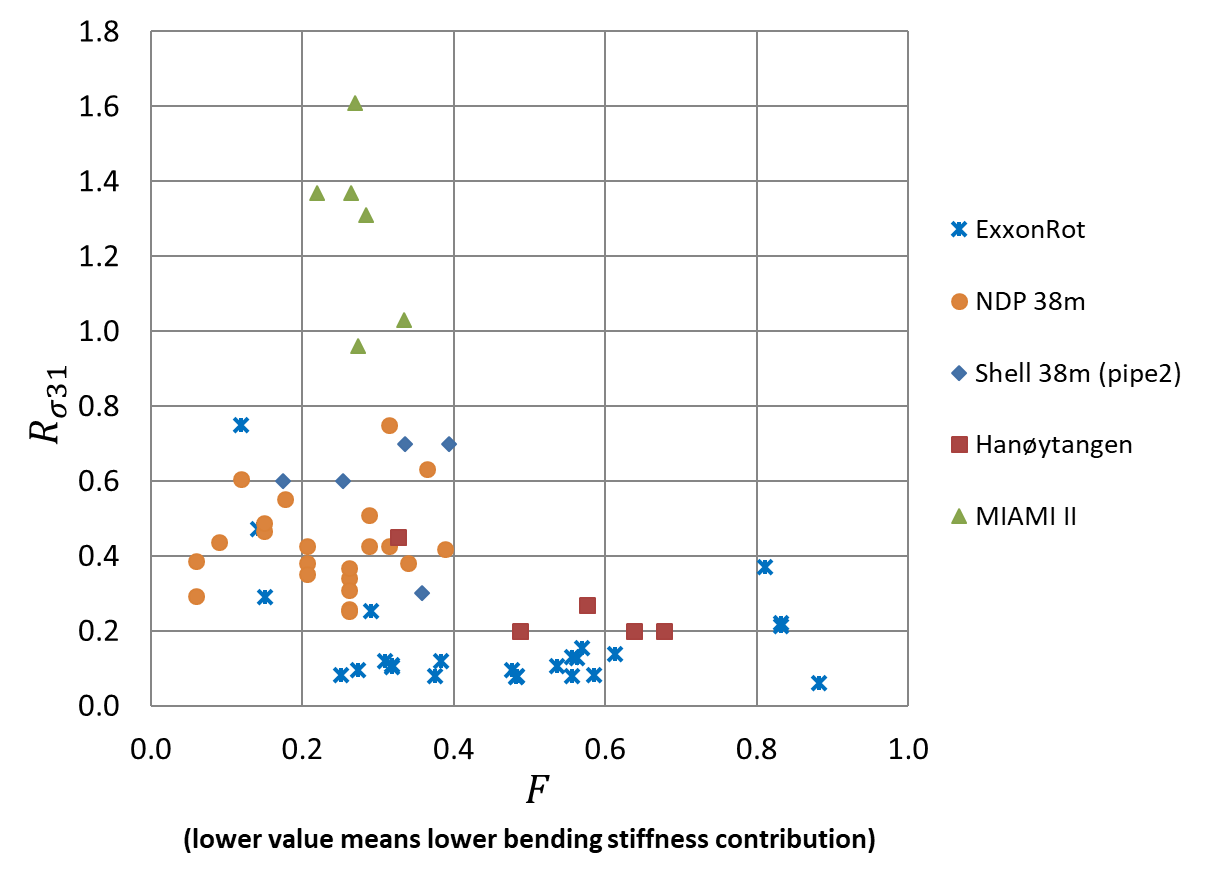}
\caption{Bending stress ratio vs. relative bending stiffness contribution. The collective sample of experiments jointly covers a large range in parameter space with each region being covered by more than one experiment apart from the stress ratios and highest relative bending stiffness contributions.}
\label{fig:3omega_tests}
\end{figure}

The characteristics of VIV responses subjected to a sheared current profile have been studied in \cite{wu2018}, as summarised in Table \ref{tab:vivc1} and in Figure \ref{fig:3omega_tests}. In Figure \ref{fig:3omega_tests}, the stress ratio ($R_{\sigma 31}$) is defined as the ratio between the standard deviation of the maximum stress measured at ($3\times \omega$) and ($1\times \omega$) frequencies along the length of the test pipe. 
\begin{equation}
    R_{\sigma 31}=\frac{max(std(\sigma_3))}{max(std(\sigma_1))}
\end{equation}
The bending stiffness ratio ($F=f_{n,b}/f_{n,tot}$) indicates how significant the bending stiffness contribution is to the total stiffness of the structure. The main observation from Figure \ref{fig:3omega_tests} is that the stress ratio increases when the bending stiffness is of less importance. In addition, the stress ratio can be very different depending on the presence of travelling waves in the response, which is indicated by the data points around $f_{n,b}/f_{n,tot}=0.3$. This study shows that there needs to be long regions of stable and favourable motion orbits inside the excitation region in order to get significant third order ($3\times \omega$) harmonic stresses. This 
seems to require the following conditions:
\begin{itemize}
    \item Response dominated by travelling waves
    \item Low relative bending stiffness contribution 
    \item Stationary response
\end{itemize}

These conditions will not only influence the higher harmonic stress, but also the responses at the primary shedding frequency ($1\times \omega$). The strong response at the primary shedding frequency is normally associated with a strong higher harmonic response \cite{wu2018}. 

\begin{table}[H]
\caption{Summary of response characteristics from selected model tests \cite{vivbp2016}.} \label{tab:vivc1}
\centering
\begin{tabular}{lcp{3cm}lp{2cm}}
\toprule
\textbf{TEST}	& \textbf{Mode order}	& \textbf{Travelling/standing wave response} & \textbf{Stiffness} & \textbf{$3^{rd}$ order harmonic stress}\\
\midrule
NDP high mode test	& 4 – 14 &	Increased travelling wave response with mode order &	Tension dominated &	Medium\\
Shell high mode test (pipe 2) &	2 – 14	& Increased travelling wave response with mode order &	Tension dominated	& Medium\\
Hanøytangen test &	10 – 27 &	Travelling waves	& Bending dominated	& Low\\
Miami II test &	25 – 39 &	Travelling waves	& Tension dominated	& High\\
ExxonMobil rotating rig test &	1 – 7 &	Standing waves	& Bending dominated &	Low\\
\bottomrule
\end{tabular}
\end{table}

\section{Limitations of the present empirical VIV prediction tools} \label{sec:limitations}
Present empirical VIV prediction tools are based on a response frequency model and a load model with a hydrodynamic coefficient database. The structural model is linear.
\subsection{Response frequency model}
In order to describe the frequency variations in space and time, the prediction tools include ``space sharing'' and/or ``time sharing'' models. In the space sharing model, the responses in different sections of the pipe may be dominated by different frequencies, but these frequencies will be constant in time. On the contrary, time sharing is assumed to appear as a sequence of single frequency responses \cite{Larsen2009}. These frequency models are reasonable approximations of the observed physical process, but with strong simplifications. There is also a lack of criteria when to switch between space sharing and time sharing models in the analysis, and consequently the analysis will be either time sharing or space sharing when it should have been a combination. 
\subsection{Hydrodynamic parameters}
Present empirical VIV prediction tools rely on hydrodynamic parameters generalised from rigid cylinder tests. The CF hydrodynamic data is based on tests with pure CF harmonic motions \cite{gopthesis}. The excitation coefficients are dependent on the motion amplitude ratio $({A/D})$ and the non-dimensional oscillation frequency ${(f_{osc}D/U)}$. However, it is known that hydrodynamic coefficients of an elastic pipe will be strongly influenced by IL motions, which are not present the in one-dimensional motion tests \cite{aronthenthesis,bourguet2013,wu2016a}. 

Inverse analysis has also been applied to estimate CF hydrodynamic coefficients of an elastic pipe from measured response data directly \cite{wu2009}.  The results show that the excitation coefficients of an elastic cylinder are different compared to those from a rigid cylinder with pure CF motions. It is also demonstrated that the coefficients are further influenced by multiple frequency variations in space and time. These frequencies compete for energy, which leads to less stable vortex shedding and more chaotic responses. Hence, the magnitudes of the excitation coefficients are smaller compared to that of a stable response dominated by one single frequency. These effects have not been fully accounted for in the existing hydrodynamic coefficients.


The hydrodynamic coefficient database can be modelled by a set of parameters. An optimisation technique has been applied to obtain best fit hydrodynamic parameters by calibrating the model prediction against measured response data from elastic cylinder tests. The space sharing model has been used in the calibration. The fatigue prediction using calibrated parameters agrees well with the measurements \cite{mukundan2010,voie2017}.

\begin{figure}
\centering
\includegraphics[width=0.8\textwidth]{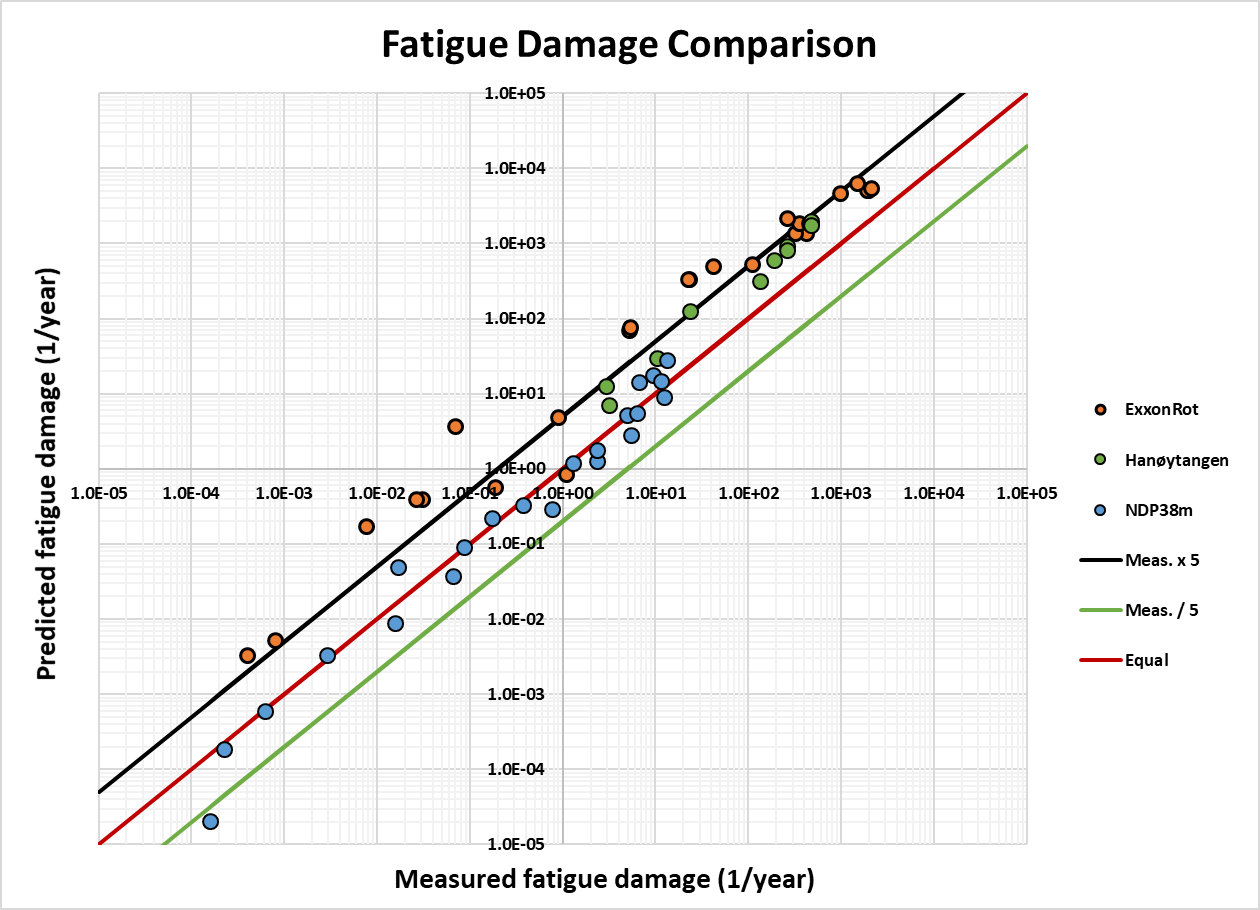}
\caption{Fatigue prediction based on one set of hydrodynamic parameters.}
\label{fig:Fatigue_original}
\end{figure}

However, increased discrepancy between prediction and measurement has been observed when the same set of hydrodynamic parameters is applied to predict VIV responses of different model tests as illustrated in Figure \ref{fig:Fatigue_original}. The ratio between the measured maximum fatigue damage and the predicted results exceeds a factor of 5 for about 26\% of the 57 cases. 53\% of the cases have prediction errors within a factor of 3. The largest deviation shows over-predictions by a factor of 52. In present study, one single slope S-N curve has been applied to all the cases in the fatigue calculation , which is presented in:

\begin{equation}
   log(N)=log(a)-mlog(S)
\end{equation}
where, N is the predicted number of cycles to failure under stress range S (in Mpa), m is the inverse slope of the S-N curve, and log used in the notation is log to the base 10. The S-N curve constant m and a is 3 and 11.63 respectively.

\section{Improved VIV prediction accuracy with adaptive hydrodynamic parameters} \label{sec:adaptive}
\subsection{Concept}
The challenge is how to compensate for the simplicity in the load model and the hydrodynamic parameters so that the prediction matches the actual VIV responses from measured data. Optimisation is one way to obtain best-fit parameters, however, it is difficult to obtain a single set of parameters which give accurate predictions for all the cases as shown in Figure \ref{fig:Fatigue_original}. The more complicated cases to include, the less robust the obtained parameters will be. The proposed strategy is therefore to apply adaptive parameters that are reflecting the different response patterns. A four-step approach is applied in the present study:
\begin{enumerate}
    \item First, the measured VIV response will be analysed to identify key parameters to represent the response characteristics. These parameters will be grouped by similarity using data clustering algorithms. 
    \item Secondly, optimal hydrodynamic parameters will be identified for each data cluster by optimisation against measured data within that cluster. 
    \item Thirdly, the VIV response using the obtained parameters will be calculated and the prediction accuracy evaluated. 
    \item Last but not the least, new cases must be assigned to a cluster in order to determine the correct hydrodynamic parameters. An iteration of the previous steps may be needed if the prediction accuracy of the new case is not satisfactory. 
\end{enumerate}

In the present study, CF VIV responses at the primary response frequency have been evaluated. The main objective of the present work is to demonstrate the validity/applicability of the concept and hence, simplifications in the analysis procedure have been made. 

\subsection{Clustering the model test data} \label{sec:clusteringdata}
\begin{figure}
\centering
\includegraphics[width=0.49\textwidth]{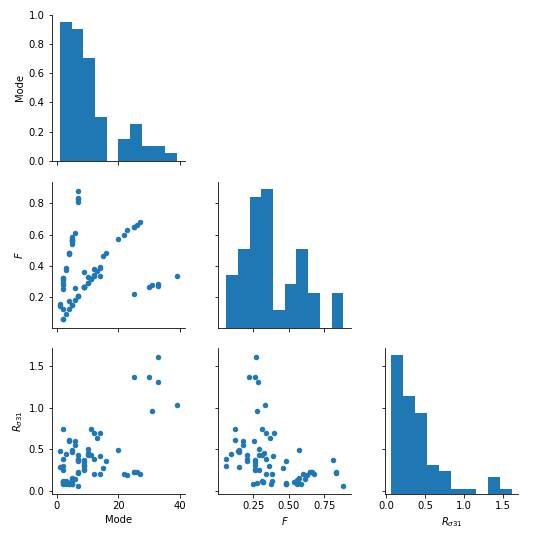}
\includegraphics[width=0.49\textwidth, trim={6cm 0 0 0}, clip]{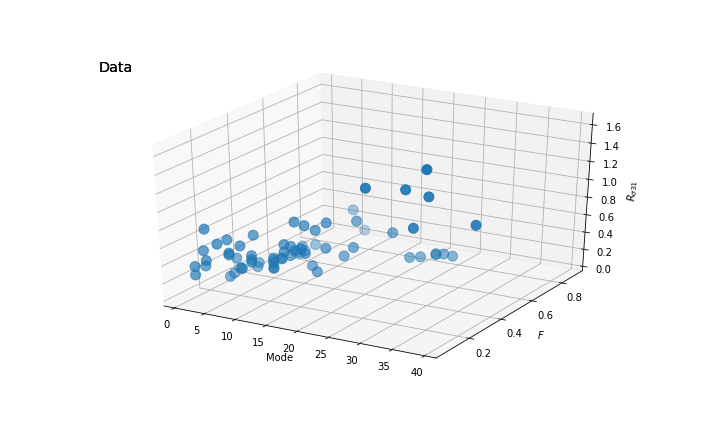}

\caption{Scatter plots and histograms for the parameters selected in Sec. \ref{sec:characteristics}: The oscillation mode, stress ratio ($R_{\sigma 31} = max(std(\sigma 3))/max(std(\sigma 1))$), and relative bending stiffness ratio ($F = f_{n,b}/f_{n,tot}$). }
\label{fig:data_scatter}
\end{figure}

Instead of using one set of generalised parameters we suggest to cluster the model tests based on the physical response properties rather than the individual model test properties, and to derive a set of parameters to be used with each cluster. The stress ratio ($R_{\sigma 31}$), bending stiffness ratio ($F$) and response mode number ($n$) are selected as inputs to the data clustering analysis, as they seem to be good indicator of the different riser response characteristics, refer to Section \ref{sec:characteristics}. The outputs are clustered model test data.

These parameters are continuous, but as seen from Figure \ref{fig:data_scatter}, they are not uniformly sampled by the model tests. We wish to identify clusters of similarity in the data under the assumption that the underlying distributions are continuous and any new model tests might close the gaps that at the moment seem to be obvious splits in the data. Consequently, we need a dynamic approach where the clustering can be easily adapted to new measurements, and also expanded to include different/additional parameters.

In recent years there has been a massive development of machine learning methods for clustering data based on similarities. Given the continuous nature of the parameters and possible overlap between clusters, conceptually different methods for clustering were tested and their performance compared. Based on the limited data sample presented here, they all perform similarly. 


In the following, the application of various clustering methods is investigated. It is expected that the underlying distributions are continuous but can be represented as Gaussian distributions with additional overlaps due to the measurement uncertainties. The method of choice should be able to handle this. The presented clustering algorithms all require a fixed number of clusters to be pre-defined. In order to have a sufficient high number of data points in the resulting clusters to derive the best fit parameters, three clusters were used. As more data become available, the number of clusters should be adjusted. Before the data can be clustered it is customary to normalise all parameter values to the same numeric scale. This has been achieved with the min-max-scaling: $x_\mathrm{i, \, scaled} = (x_i - x_{\min})/(x_{\max} - x_{\min})$.

\begin{figure}[!h]
\centering
\includegraphics[width=0.49\textwidth]{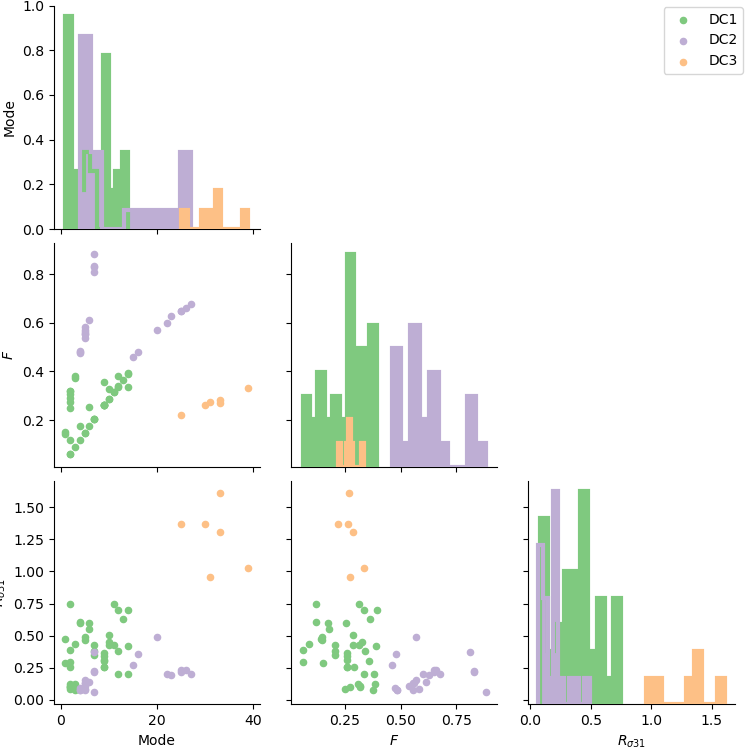}
\caption{Scatter plots for the resulting three clusters obtained for the K-Means algorithm in the parameter space defined by the oscillation mode, stress ratio ($R_{\sigma 31}=max(std(\sigma_3))/max(std(\sigma_1))$), and relative bending stiffness ratio ($F = f_{n,b}/f_{n,tot}$). The method assumes the clusters to be spherical in parameter space.}
\label{fig:kmeans}
\end{figure}

{\bf K means clustering} \cite{lloyd1982least} is a standard algorithm often used because of its simplicity. However, it does not cluster the data based on similarities, but it is partitioning the data into a user-defined number of chunks assumed to be spherical in the parameter space. This works well when the data consist of a known number of clearly separated clusters. However, the method is not very useful for deriving the number of clusters in a scenario with possible overlaps. The results from K means clustering is shown in Figure \ref{fig:kmeans}.

\begin{figure}[!h]
\centering
\includegraphics[width=0.49\textwidth]{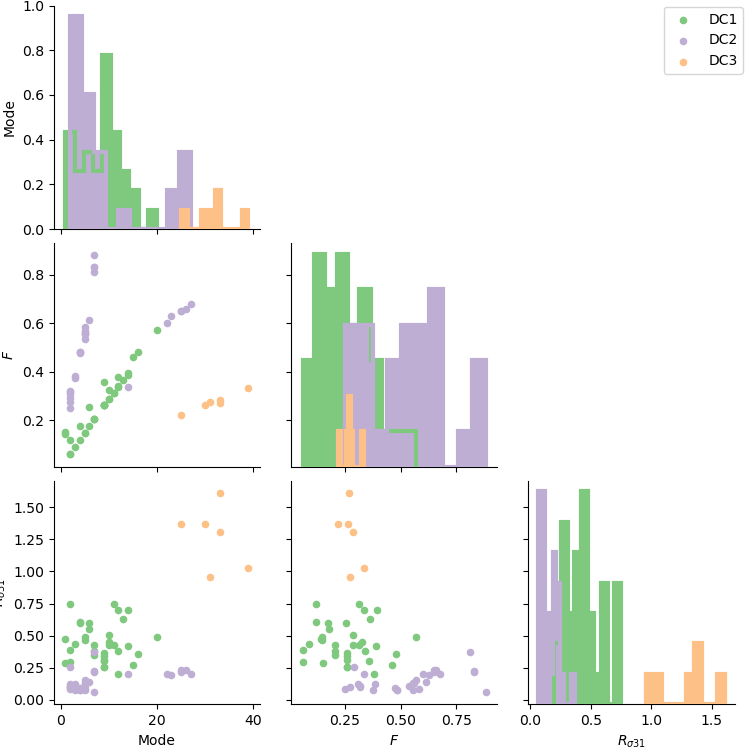}
\caption{The resulting three clusters from the Gaussian mixture algorithm in the same parameter space as for Figure  \ref{fig:kmeans}. The clusters are assumed to be overlapping and non-symmetric Gaussians in the parameter space. The distributions of the individual model tests among the clusters are given in Table \ref{tab:cluster}.}
\label{fig:gaus}
\end{figure}

{\bf Gaussian Mixture} \cite{bishop2006pattern, attias2000variational, blei2006variational} is a probabilistic model that assumes that all data points originate from a pre-determined number of overlapping Gaussian distributions with unknown parameters. The results of Gaussian mixture clustering is shown in Figure \ref{fig:gaus}.

\begin{figure}
\centering
\includegraphics[width=0.49\textwidth]{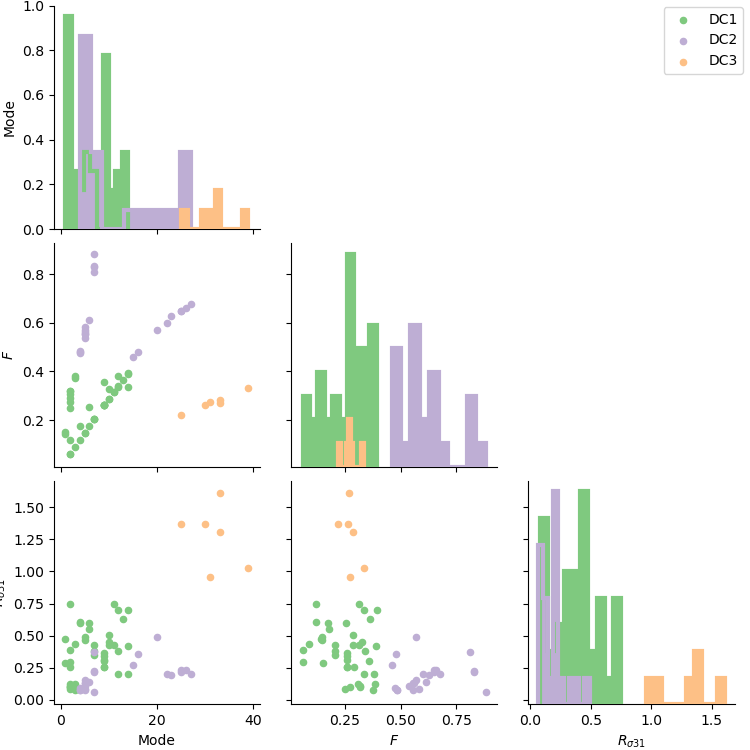}

\caption{The result of spectral clustering for three clusters in the same parameter space as for Figure \ref{fig:kmeans}. The three algorithms agree on the small high-mode cluster, and also on the main structure of the remaining two clusters, but not the exact distribution of the data points. The Gaussian mixture allows for larger overlap between the clusters than the spectral clustering.}
\label{fig:spec}
\end{figure}

{\bf Spectral clustering} \cite{Shi00normalizedcuts, Luxburg07atutorial} uses the spectrum (eigenvalues) of the similarity matrix of the data to reduce the number of dimensions before clustering in the low-dimension space. The number of clusters has to be specified and it only works for a low number of clusters. The results of spectral clustering is shown in Figure \ref{fig:spec}.

As seen in Figures \ref{fig:kmeans}-\ref{fig:spec}, the different data clustering methods give very similar results for the parameter clustering, but they are not identical. The robustness of the clusters should be investigated with more data. The subsequent analysis was performed with the Gaussian mixture algorithm since the data are assumed to originate from continuous distributions of physical origin and can therefore be reasonably approximated with overlapping Gaussian distributions. 


\subsection{Identifying response characteristics by data clustering}
In the present study, VIV response characteristics are represented by three parameters based on the five selected test data sets. The number of selected data points from various model tests and their allocations in three data clusters (DC 1-3) are reported in Table \ref{tab:cluster}. 

\begin{table}[H]
\caption{Results of Gaussian mixture clustering analysis.} \label{tab:cluster}
\centering
\begin{tabular}{lccccc}
\toprule
\textbf{Test}	& \textbf{Number of cases}	& \textbf{DC1}  & \textbf{DC2} & \textbf{DC3}  \\
\midrule
NDP high mode test	&	22&	22&	0 & 0 \\
Shell high mode test (pipe 2) 	&	6&	5&	1 &	0 	\\
Hanøytangen		&	11& 5	& 6	& 0		\\
ExxonMobil rotating rig test		&	25&	3& 22	&	0 	\\
Miami II test 			&	6&	0 &	0&	6 	\\
\bottomrule
\end{tabular}
\end{table}

It can be seen from Table \ref{tab:cluster} that the data from the Hanøytangen test has been split into two roughly evenly sized groups associated with data clusters DC1 and DC2, respectively. In the Hanøytangen test, the structural stiffness for cases with lower response mode (lower towing speeds) were tension dominated. The response characteristics in terms of response magnitude and frequency contents are similar to NDP and Shell tests. But, the bending stiffness becomes dominant with increasing mode number. The response characteristics tend to be similar to the ExxonMobil test. The results from the clustering analysis agree with observations from the model tests. This illustrates the purpose of clustering analysis, which groups the cases with the same characteristics across different model tests. It is also shown in Figure \ref{fig:3omega_tests} that it is difficult to use the bending stiffness ratio as one single parameter to cluster the data. The data from MIAMI showed significantly different behaviour as compared to other data with a similar bending stiffness ratio about 0.3. The travelling speeds in CF and IL directions will be close due to its relatively low bending stiffness. In addition, the MIAMI model is responding at a very high mode, where the travelling wave response is dominating. Therefore, there is a long region along the length of the riser model where stable and favourable motion orbits are observed, which leads to high third order stress as well as to high responses at the fundamental frequency. On the other hand, the other cases with similar or lower bending stiffness ratio have much lower the third order stress due to that the motion orbits vary constantly along the length, which is caused by the presence of the standing waves at low response modes. The number of data groups is manually set to be 3 based on the data at hand. The clustering analysis correctly identified three clusters as requested. An optimal selection of the clusters will be expected when dealing with more complex data.

\subsection{Obtaining optimal hydrodynamic parameters} 
The excitation coefficient database is represented by a few parameters \cite{gopthesis}. They are systematically varied until the prediction agrees with the measurement as explained in \cite{voie2017}. The optimisation is carried out for each of the three data clusters. The mean squared error in fatigue damage has been used as the criterion to select the optimal parameters, i.e. $\sum_{i=1}^{N}(Fat_{pred,i}-Fat_{meas,i})^2/N $, where $i=1,2,3…N$, $i$ is the sensor number along the length of the pipe. Two sets of hydrodynamic parameters have been derived, i.e., Ce1 and Ce2. Parameter set Ce1 was used to predict fatigue damage for cases in data cluster 1 (DC1) and the other set was used for data cluster 2 (DC2). No hydrodynamic parameters have been derived for data cluster 3 (DC3) due to lack of experimental data time series from which to derive the parameters and to evaluate the fatigue damage. However, it is expected that a third set of hydrodynamic parameters is needed because that the averaged displacement standard deviation of DC3 is indeed larger than that of the other data groups \cite{SwVa08}.  

\begin{figure}
     \centering
     \begin{subfigure}[b]{0.48\textwidth}
         \centering
         \includegraphics[width=\textwidth]{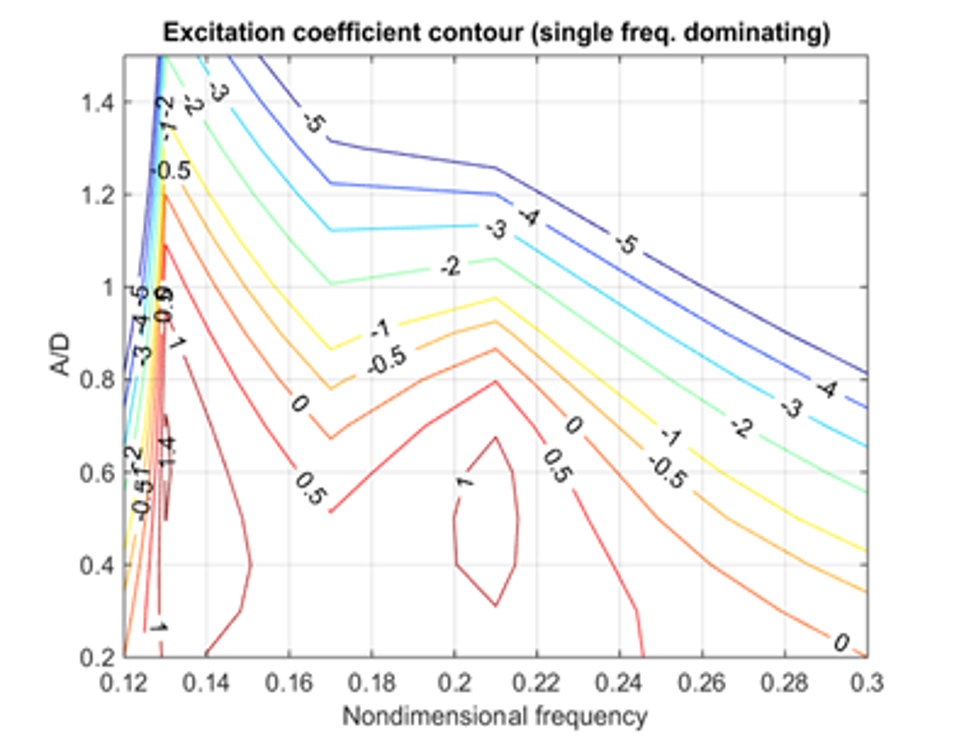}
         \caption{Single-frequency response coefficients (Ce1).}
         \label{fig:ce_single}
     \end{subfigure}
     \hfill
     \begin{subfigure}[b]{0.48\textwidth}
         \centering
         \includegraphics[width=\textwidth]{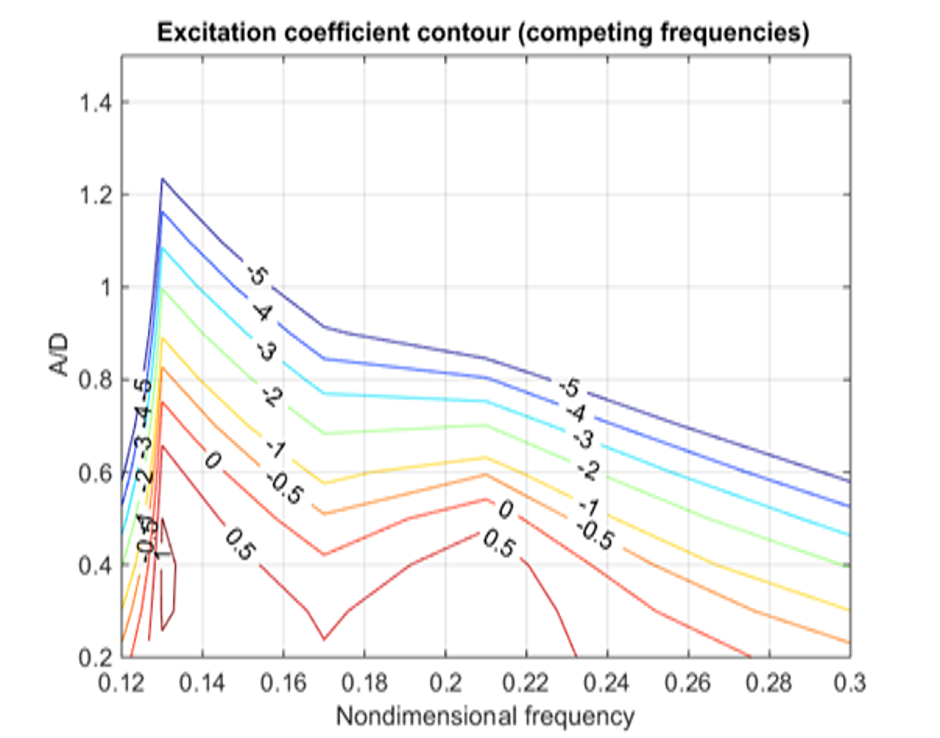}
         \caption{Multi-frequency response coefficients (Ce2).}
         \label{fig:ce_multi}
     \end{subfigure}
        \caption{CF excitation coefficient contour plot corresponding to two sets of hydrodynamic parameters.}
        \label{fig:ce_adapt}
\end{figure}

The reconstructed CF excitation coefficient contour plots are presented in Figure \ref{fig:ce_adapt}. The first set of parameters (Ce1) as represented by the force coefficient contour plot to the left were developed in earlier studies \cite{voie2017}. This is representative for a tension stiffness dominated pipe dominated by a single response frequency. The second set of hydrodynamic parameters (Ce2), as shown in the plot to the right, is adjusted to represent the competing frequency responses of a bending stiffness dominated pipe.

\subsection{Evaluating prediction accuracy}
\begin{figure}
\centering
\includegraphics[width=0.8\textwidth]{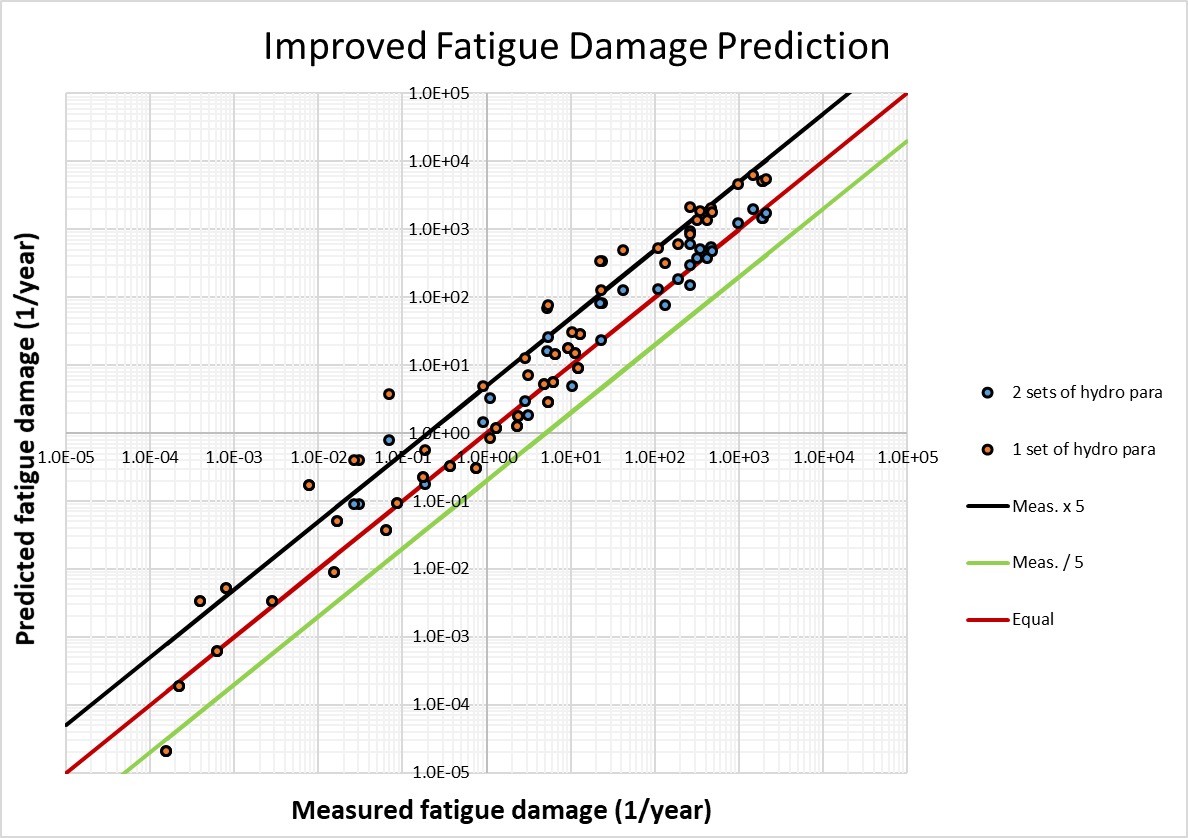}
\caption{Comparison of the predicted fatigue relative to the measured fatigue when predicted using one or two optimal sets of hydrodynamic parameters. The deviation between the predictions and measurements is clearly smaller when using two sets with the allocation of parameters based on the clustering analysis.}
\label{fig:Fatigue_opitmal}
\end{figure}

In Figure \ref{fig:Fatigue_opitmal}, the fatigue damage predictions of those cases, for which a calculation with two sets of hydrodynamic parameters has been performed, are compared to the respective predictions using only one parameter set. The red line indicates that the predicted fatigue damage is equal to the measured fatigue damage. The black lines shows that the predicted fatigue damage is 5 times of the measured value, and the green line shows that the predicted fatigue damage is 1/5 of the measured value. The narrower distribution along the diagonal obtained when using two sets of hydrodynamic parameters indicates a significant improvement in prediction accuracy. Only 7\% of the predictions  deviate by more than a factor of 5 from the measured fatigue damage, compared to 26\% when using only one set of parameters. 86\% of the cases have prediction error within a factor of 3, which represents an increase of 33\% compared to the previous analysis with only one set of parameters. 

\subsection{Selecting of hydrodynamic parameters for a new data by classification algorithms}
When new measurement data becomes available, the clustering model  can be applied to determine which parameter sets should be used. Assuming a new test case with the following conditions: mode number of 25, bending stiffness ratio of 0.6 and stress ratio of 0.1. The developed Gaussian mixture model in Sec. \ref{sec:clusteringdata} determines that this new data point will be classified to DC2, and consequently the best fitting parameters for DC2 should be applied to compute the fatigue. 

The prediction accuracy needs to be evaluated and data re-clustering may be needed if the prediction accuracy is not satisfactory. New parameters may be identified and included in a new data clustering procedure. For example, Reynolds number may be an additional clustering parameter when dealing with full scale measurements \cite{lie2013,yin2018b}.

\section{Discussions} \label{sec:discussion}
The novelty of the present work is to apply adaptive parameters in the prediction in order to account for the limitations of the simplified numerical models.The key element is to group the VIV responses and apply different hydrodynamic parameters for each individual response cluster. It is found increasingly difficult to derive analytical functions to describe the relationship among the increasing number of parameters. It is also not desired to manually inspect the data, especially when the accumulation of data is large. Therefore, data clustering methods have been explored for grouping data in a multi-dimensional parameter space in the present study. For the demonstration purposes, the strategy in the present work is to limit the scope to a relatively small data group as a first step. The data has been limited to selected laboratory tests with dense instrumentation with sheared current profiles. The good physical understanding of these limited cases provides a way to evaluate the use of the clustering analysis, which groups the data purely based on their similarities without physical meaning. three parameters, i.e., stress ratio, stiffness ratio and mode order (refer to Figure \ref{fig:3omega_tests}), were selected based on the physical understanding of the system. It is satisfactory to see that the clustering analysis lead to a similar selection of groups as otherwise could have been achieved manually. However, the ultimate goal is to eventually apply this concept for VIV predictions in riser design practice and for VIV field response monitoring, where increasing complexity in data quantity, quality and physical systems are expected. Therefore, a more advanced clustering analysis is to be expected in practice.

More model test data need to be included in the analysis to increase the robustness of the parameter selection. More parameters are also needed in order to describe the characteristics of the riser response, e.g., the shape and direction of the flow profile, Reynolds number effect, etc. The clustering analysis will hopefully help to identify the in-direct connections in the data in a multi-parameter space. Then, the improvement in modelling will be easier with the clear understanding of the limitation of the prediction tools. This process is also expected to be dynamic, which means that the improvement in the numerical models may potentially reduce the number of parameter sets otherwise required. This will be carried out step-wisely so that the underlying physics are correctly understood.

The proposed concept can be applied to improve accuracy in the riser design and in on-line response monitoring. While stiffness ratio and even mode order can be predicted, stress ratio cannot be estimated beforehand. Eventually, other parameters will be used.

The geometry, operational and environmental conditions of the riser systems in the field are much more complicated than in the laboratory tests. In addition, the field measurement quality and limited sensor density will introduce significant uncertainties. Field measurement data must be analysed to identify key parameters to describe response characteristics. These parameters and their values are not expected to be identical to those used in the design phase. But, the knowledge gained from the analysis of laboratory data will provide a good basis for parameter selection and understanding of the field observations. Learning from field measurements will certainly be valuable to improve the design basis as well.

%
%
\section{Conclusions}
A new analysis concept to improve VIV prediction accuracy by applying adaptive parameters has been proposed and demonstrated with simplified examples. The limitations of the mathematical model and the hydrodynamic parameters used in the existing perdition tools can be compensated by parameter optimisation based on groupings of the measured data. The identification and selection of these groups were determined adaptively by using data clustering algorithms. Improvement of both the prediction accuracy and consistency have been achieved. By using the adaptive parameter approach, 93\% of the predictions are within a factor of 5 of the measured values, and 86\% of the predictions are within a factor of 3 of the measured values. The corresponding percentages are 74\% and 53\% using original parameters.


The proposed analysis concept is semi-empirical, which means it is based on model test data. Enriching model test data will help increase the robustness of the parameters, so that the clustering analysis and VIV prediction will be improved. It is expected that the proposed analysis concept can  be applied in both riser design and on-line riser response monitoring applications.

\section{Acknowledgement}
The authors would like to thank Equinor, BP,
Kongsberg Maritime, Trelleborg Offshore, Aker Solutions and Subsea7 for funding and permission to publish this study. The authors would like to thank Professor Emeritus Carl M. Larsen and Dr. Elizabeth Passano for fruitful discussions.

%
%
%
%
\authorcontributions{Conceptualization, J.W.; methodology, J.W., S. R.-S., D.Y. H.L. and S.S.; writing--original draft preparation, D.Y., J.W. S. R.-S.; writing--review and editing, H.L., S.S. and M.T.}

%
\funding{This research is funded by the joint industry project - Pilot PRAI, funded by SINTEF, Equinor, BP, Kongsberg Maritime, Trelleborg Offshore, Aker Solutions and Subsea7.}
%
%
\conflictsofinterest{The authors declare no conflict of interest.}
\reftitle{References}


\externalbibliography{yes}
\bibliography{mdpi-jmse_pilot_paper1}



\clearpage
\newpage
\section*{Appendix: Description of different model tests}
\subsection*{NDP high mode VIV test}
In December 2003, Norwegian Deepwater Programme (NDP) carried out a VIV test programme in the MARINTEK's Ocean Basin focusing on conditions leading to high vibration mode numbers \cite{trim2005}. A 38 m long elastic pipe model with an outer diameter of 0.027 m was towed at various speeds, simulating both uniform and linear sheared currents. The two ends of the test model were pinned and pre-tensioned. Bending strains and accelerations along the model were measured. Both a bare riser configuration and configurations with two types of strakes for VIV suppression were tested. The Reynolds number ranged from 8000 to 67500. The calculated bending stiffness ratio indicated that bending stiffness of the pipe was dominated by the geometrical stiffness from tension ($f_{n,b}/f_{n,tot}<0.5$) and the bending stiffness was found to be less important. For the low vibration mode cases ($n <10$), the response frequency was dominated by a single frequency. For the higher mode cases, response amplitudes, dominating frequency and mode composition were seen to vary in time. Travelling wave responses were clearly observed at the highest response modes (mode 10-14) \cite{trim2005,wu2018}.

\subsection*{Shell high mode VIV test}
Shell performed high mode VIV tests in the Ocean Basin at MARINTEK in 2011 \cite{HaHe12}. A test setup similar to the NDP high mode VIV test was applied. Three different heavily instrumented riser models were towed at different speeds, simulating uniform and linearly shear currents. The highest Reynolds number reached was $ 2\times 10^5 $. Measurements of micro bending strains and of accelerations along the risers were made in the IL and CF directions. 
The results from one of the riser models, Pipe 2, are used in the present study. Pipe 2 was 38 m in length and 3 cm in diameter. The response characteristics are similar to the NDP test.

\subsection*{Hanøytangen high mode VIV test}
The Hanøytangen VIV test was carried out in a fjord outside of Bergen in Norway in 1997 \cite{otc8071}. The purpose of the Hanøytangen test program was to create data for studying the VIV response of deep sea risers subjected to sheared currents. The riser model was suspended from a catamaran type surface vessel, and the lower ends were attached to buoyancy tanks by ropes via pulleys. The buoyancy tanks provided the pre-tension. The catamaran was towed along the quay by means of a closed rope loop running over pulleys. The length of the steel pipe model was 90 m and the diameter was 3 cm. The dynamic responses became dominated by the bending effects with increasing response mode orders (mode 10-27). In addition, resonant axial vibrations were observed, which may lead to significant stresses.

\subsection*{ExxonMobil rotating rig VIV test}
ExxonMobil performed a VIV test on a long elastic cylinder \cite{Tognarelli2004} at MARINTEK's towing tank. A rotating test rig was used to expose a 9.63 m long test pipe with 20 mm diameter to flow conditions simulating uniform and linearly sheared currents. Both bare and straked pipes were tested. The model was heavily instrumented with bending strain gauges and accelerometers. The test model was made of brass and the pipe stiffness was dominated by the bending stiffness. Low response mode (1-7) with standing wave responses were mostly observed.

\subsection*{Deepstar Miami II test}
Field measurements of an elastic pipe were carried out by MIT and the Deepstar Consortium, known as the Miami II test \cite{Vandiver2009}. A fiber glass composite pipe with a length of 152.4 m and an outer diameter of 3.63 cm was tested. The top end of the pipe model was attached to a boat, which was put on various headings relative to the Gulf Stream so as to model a large variety of currents, varying from nearly uniform to highly sheared in speed and direction. A railroad wheel was attached to the bottom of the pipe to provide tension. The stiffness of the pipe model was tension dominated even at high mode numbers ($n >20$). IL and CF responses are likely to travel at the same speed along the cylinder. Long regions of travelling waves were observed in the tests.
\end{document}